\documentstyle[12pt]{article}
\newlength{\myleftmargin}
\newlength{\paperwidth}
\newcommand{\mq}{1/m^{}_Q}
\newcommand{\mql}{\frac{\lambda_1}{2m^{}_Q}}
\newcommand{\mqll}{\frac{\lambda_2}{2m^{}_Q}}
\newcommand{\mqlt}{\frac{\lambda_1^{0^+}}{2m^{}_Q}}
\newcommand{\mqls}{\frac{\lambda_1^{1^+}}{2m^{}_Q}}
\newcommand{\mqlls}{\frac{\lambda_2^{1^+}}{2m^{}_Q}}
\newcommand{\slp}{s^P_{\ell}}
\newcommand{\cc}{\cos^2\theta}
\newcommand{\sss}{\sin^2\theta}

\begin{document}

\thispagestyle{empty}

\begin{flushright}
DPNU-96-26 \\
May 1996
\end{flushright}

\vspace{2em}

\begin{center}
{\Large{\bf The $\Xi_Q$-$\Xi_Q'$ Mixing and Heavy Baryon Masses}}

\vspace{5em}

{{\sc Toshiaki Ito} and {\sc Yoshimitsu Matsui}}

\vspace{2em}

\sl{Department of Physics, Nagoya University,}~ 
{\it Nagoya 464-01, Japan}
\end{center}

\vspace{5em}

{\large {\bf Abstract :}}~~
We examine the ${\sl\Xi}_Q$-${\sl\Xi}'_Q$ mixing
and heavy baryon masses in the heavy quark effective theory with the
$\mq$ corrections.  In the conventional baryon assignment, we obtain
the mixing angle $\cos ^2 \theta=0.88\pm 0.12$ in virtue of the
Gell-Mann$-$Okubo mass relation.  On the other hand, if we adopt the
new baryon assignment given by Falk, the allowed region of the
${\sl\Sigma}_c$ mass is upper from 2372MeV.

\clearpage

The heavy quark effective theory(HQET) is an efficient approach
to investigating the properties of hadrons
containing a single heavy quark $Q$.\cite{rf:1,rf:2} \ 
Heavy hadron mass spectra have been studied in the HQET
by taking into account $\mq$ corrections.\cite{rf:5} \ 
In the heavy quark symmetry limit,
the total spin of the heavy hadron $J$ and
the spin of the light degrees of freedom $s_{\ell}$ ({\it i.e.}, the angular
momentum of the light degrees of freedom in the heavy quark rest frame) are
separately conserved under strong interactions,
because the heavy quark spin $s_Q$ decouples from the hadron dynamics.
The hadron containing a heavy quark is the simultaneous eigenstate of the 
energy, the total angular momentum-parity $J^P$ and the spin-parity of
the light degrees of freedom $s_{\ell}^P$.
Therefore, the heavy hadrons can be classified by $s_{\ell}^P$
rather than the relative internal orbital angular momentum.\cite{rf:6} \ 
At the infinite quark mass limit({\it i.e.}, $\Lambda_{QCD}/m_Q\to 0$),
the states with the same $s_{\ell}^P$ become degenerate.
By incorporating the effect of the $1/m_Q$ correction,
this degeneracy is resolved.
In this case the mixing between two states
with the same total spin-parity is unavoidable in general
since they are no longer eigenstates of $\slp$.
For instance,
${\sl\Xi}_Q$ and ${\sl\Xi}'_Q$ mix with each other.
In this paper,
we discuss the ${\sl\Xi}_Q$-${\sl\Xi}'_Q$ mixing and heavy baryon mass spectra
in both the cases of the conventional assignment and the new assignment
given by Falk.\cite{rf:7}

The HQET Lagrangian with $\mq$ corrections is
\begin{equation}
{\cal L} = {\bar h}_v(iv\cdot D)h_v+\frac{1}{2m^{}_Q}{\bar h}_v(iD)^2h_v
+Z_Q\frac{1}{2m^{}_Q}{\bar h}_v\sigma^{\mu\nu}G_{\mu\nu}h_v,
\end{equation}
where $v$ is the four velocity of the heavy quark $h$,
$Z_Q$ is a renormalization factor and $G_{\mu\nu}$
is the gluon field strength.\cite{rf:8}
Neglecting higher order corrections,
we have a heavy hadron mass formula
\begin{equation}
M_{H_Q} = m^{}_Q+{\bar\Lambda}- \mql -d^{}_H\mqll, \label{eqn:mass}
\end{equation}
where ${\bar\Lambda}$ is the energy of the light degrees of freedom
in the hadron.
Two $\mq$ correction coefficients are given by the matrix elements
\begin{equation}
\lambda_1 = \langle H_Q(v)|{\bar h}_v(iD)^2h_v|H_Q(v)\rangle
\end{equation}
and
\begin{equation}
d^{}_H\lambda_2 = \frac{1}{2}Z_Q\langle H_Q(v) |
{\bar h}_vG_{\mu\nu}\sigma^{\mu\nu}h_v |H_Q(v)\rangle ,
\end{equation}
where $d^{}_H$ is the Clebsch-Gordan factor.

Let us apply the heavy hadron mass formula (\ref{eqn:mass}) to the
ground state baryons containing a single heavy quark, $Qqq$($q=u$ or $d$).
There are three states in the ground state $Qqq$ baryons ({\it i.e.},
${\sl\Lambda}_Q$, ${\sl\Sigma}_Q$ and ${\sl\Sigma}_Q^*$).
The ${\sl\Lambda}_Q$ baryons have
$s_{\ell}^P=0^+$ and $I=0$,
while the ${\sl\Sigma}_Q$ and ${\sl\Sigma}_Q^*$ baryons have
$s_{\ell}^P=1^+$ and $I=1$.
The baryon with
$s_{\ell}^P=0^+(1^+)$ and $I=1(0)$
is not allowed due to the antisymmetrization of
the wave function of the baryon.
These baryon masses are given by
\begin{eqnarray}
{\sl \Lambda}_Q &=& m^{}_Q+{\bar \Lambda}_{0^+}-\mqlt, \\
{\sl \Sigma}_Q &=& m^{}_Q+{\bar \Lambda}_{1^+}-\mqls -4\mqlls, \\
{\sl \Sigma}_Q^* &=& m^{}_Q+{\bar \Lambda}_{1^+}-\mqls +2\mqlls.
\end{eqnarray}
In this paper,
the symbols which represent the baryon states
in equations stand for their masses.

If two light quarks are both $s$ quarks,
there are two baryon states ({\it i.e.}, ${\sl\Omega}_Q$
and ${\sl\Omega}_Q^*$) with $\slp =1^+$.
The $Qss$ baryon with $\slp =0^+$ does not exist for
the same reason as the case of the $Qqq$ baryons.
So the masses of ${\sl\Omega}_Q^{(*)}$ baryons are written as
\begin{eqnarray}
{\sl \Omega}_Q &=& m^{}_Q+{\bar \Lambda}''_{1^+}-\mqls -4\mqlls, \\
{\sl \Omega}_Q^* &=& m^{}_Q+{\bar \Lambda}''_{1^+}-\mqls +2\mqlls .
\end{eqnarray}

Attention should be paid
when we apply the mass formula to the ${\sl\Xi}_Q$, ${\sl\Xi}'_Q$
and ${\sl\Xi}_Q^*$
baryons whose valence quarks are $q$, $s$ and $Q$.
In the heavy quark limit, 
the spin-parity of the light degrees of freedom of these baryons is
$\slp =0^+$ or $\slp =1^+$.
The $\slp =0^+$ state has one baryon state ${\sl\Xi}_Q$ with total
spin-parity $J^P=\frac{1}{2}^+$,
while the $\slp =1^+$ state has two baryon states ${\sl\Xi}'_Q$ with
$J^P=\frac{1}{2}^+$ and ${\sl\Xi}^*_Q$ with $J^P=\frac{3}{2}^+$.
Since in the heavy quark limit the spin-parity of the light degrees of
freedom is conserved under the strong interaction,
a hadron containing a heavy quark is an eigenstate of the spin-parity
of the light degrees of freedom.
Thus ${\sl\Xi}_Q$ and ${\sl\Xi}'_Q$ never mix with each other
in this symmetry limit.
If the $\mq$ corrections are introduced,
the heavy quark spin symmetry is broken.
Then one should take the
${\sl\Xi}_Q$-${\sl\Xi}'_Q$ mixing into account.

The relation between the heavy quark limit ${\sl\Xi}_Q^H$(${{\sl\Xi}'}^H_Q$)
states and physical ${\sl\Xi}_Q$(${\sl\Xi}'_Q$)
states can be written in terms of the mixing angle $\theta$
\begin{equation}
\left(
\begin{array}{c}
|{\sl\Xi}_Q \rangle \\
|{\sl\Xi}'_Q \rangle
\end{array}
\right)
=
\left(
\begin{array}{cc}
\cos \theta & \sin \theta \\
-\sin \theta & \cos \theta
\end{array}
\right)
\left(
\begin{array}{c}
|{\sl\Xi}_Q^H \rangle \\
|{{\sl\Xi}'}^H_Q \rangle
\end{array}
\right).
\end{equation}
We may assume the mixing angle $\theta$ to be a function of
$\Lambda_{\rm QCD}/m^{}_Q$ and that it becomes zero at $m^{}_Q\to \infty$.
So the masses of ${\sl\Xi}_Q$, ${\sl\Xi}'_Q$ and
${\sl\Xi}_Q^*$ baryons are given by
\begin{eqnarray}
{\sl\Xi}_Q &=& m^{}_Q +\cc{\bar \Lambda}'_{0^+} +\sss{\bar \Lambda}'_{1^+}
 -\cc\mqlt
-\sss\mqls -4\sss\mqlls, \nonumber \\
& & \label{eqn:xiQ}\\
{\sl\Xi}'_Q &=& m^{}_Q +\sss{\bar \Lambda}'_{0^+} +\cc{\bar \Lambda}'_{1^+}
 -\sss\mqlt
-\cc\mqls -4\cc\mqlls, \nonumber \\
& & \\
{\sl\Xi}_Q^* &=& m^{}_Q + {\bar \Lambda}'_{1^+} -\mqls +2\mqlls.
\label{eqn:xiqs}
\end{eqnarray}
The mixing angle can be expressed in terms of the baryon masses,
\begin{equation}
\cc = \frac{{\sl\Xi}_Q+{\sl\Sigma}_Q^*-{\sl\Sigma}_Q-{\sl\Xi}_Q^*}
{{\sl\Xi}_Q+{\sl\Xi}'_Q+2{\sl\Sigma}_Q^*-2{\sl\Sigma}_Q
-2{\sl\Xi}_Q^*}. \label{eqn:angle}
\end{equation}
This sum rule is obtained by use of Eqs. (\ref{eqn:xiQ}) $\sim$ 
(\ref{eqn:xiqs}). 
Since heavy quark mass corrections in these equations 
are of order  $1/m^{}_Q$,
the following discussions are valid.

In this paper, we treat charmed baryons because among bottom baryons
only the ${\sl\Lambda}_b$ baryon is confirmed experimentally.\cite{rf:9} \ 
First we discuss the case of the conventional assignment for
the heavy baryons,
in which measured charmed baryon masses are assigned as
\begin{eqnarray*} 
{\sl\Lambda}_c &=& 2285\pm 1{\rm MeV},^{7)} ~~
{\sl\Sigma}_c = 2453\pm 1{\rm MeV},^{7)} ~~
{\sl\Sigma}^*_c = 2530\pm 7{\rm MeV},^{7)} \\
{\sl\Xi}_c &=& 2468\pm 1{\rm MeV},^{7)} ~~ 
{\sl\Xi}'_c = 2563\pm 15{\rm MeV},^{8)} ~~ 
{\sl\Xi}^*_c = 2644\pm 2{\rm MeV},^{9)} \\
{\sl\Omega}_c &=& 2704\pm 4{\rm MeV}.^{7)}
\end{eqnarray*}
 From Eq.(\ref{eqn:angle}) the mixing angle becomes
\begin{equation}
\cc = 0.96\pm 0.23. \label{eqn:cosin}
\end{equation}
At present, it is difficult to
determine the mixing angle definitely due to 
large experimental errors.
Here we introduce an additional condition,
\begin{equation}
{\bar\Lambda}''_{1^+}-{\bar\Lambda}'_{1^+} = {\bar\Lambda}'_{1^+}
-{\bar\Lambda}_{1^+}, \label{eqn:gell}
\end{equation}
the so-called Gell-Mann$-$Okubo mass relation.
This implies that
\begin{equation}
\frac{1}{2}({\sl\Omega}_Q-{\sl\Sigma}_Q) =
{\sl\Xi}^*_Q-{\sl\Sigma}^*_Q. \label{eqn:gellh}
\end{equation}
Substituting the measured mass values into Eq.(\ref{eqn:gellh}),
we find that
the left-hand side of this equation is $124\pm 3$MeV while the right-hand side
is $114\pm 9$MeV.
Thus this relation is consistent with the data.
By virtue of the relation (\ref{eqn:gell}), we can obtain the sum rule for
heavy baryon masses
\begin{equation}
{\sl\Xi}'_Q-{\sl\Xi}_Q = (\cc -\sss )\left(
{\sl\Sigma}_Q-({\sl\Xi}_Q+{\sl\Xi}'_Q)+{\sl\Omega}_Q\right). \label{eqn:rule}
\end{equation}
The left-hand side of this equation is $95\pm 16$MeV.
On the other hand, the second factor of the right-hand side
is $127\pm 21$MeV.
This leads to the value of the mixing angle
\begin{equation}
\cc = 0.88 \pm 0.12,
\end{equation}
which is consistent with (\ref{eqn:cosin}).
In order to determine the value of the mixing angle,
an accurate measurement of the ${\sl\Xi}'_c$ mass is needed. 

\renewcommand{\thefootnote}{\fnsymbol{footnote}}
Next let us turn our attention to Falk's assignment in which the
observed ${\sl\Sigma}_c(2453)$ is regarded as the ${\sl\Sigma}_c^*$
 state.\cite{rf:7} \ 
The advantage of this assignment is the fact that we can solve
the ${\sl\Sigma}_b$-${\sl\Sigma}^*_b$ mass problem\footnote[2]{We can obtain
the relation
\begin{displaymath}
\frac{{\sl\Sigma}^*_b-{\sl\Sigma}_b}{{\sl\Sigma}^*_c-{\sl\Sigma}_c}=
\frac{B^*-B}{D^*-D}
\end{displaymath}
 from the mass formula (\ref{eqn:mass}).
However, the masses of ${\sl\Sigma}_b$ and ${\sl\Sigma}^*_b$ measured by
the DELPHI Collaboration\cite{rf:12} are not consistent with this relation.
This is a serious question for the HQET.}.
In this assignment, the ${\sl\Sigma}_c$ state is unknown.
 From Eq.(\ref{eqn:angle}),
we obtain
\begin{equation}
{\sl\Sigma}_c > 2372{\rm MeV}.
\end{equation}

Here it is useful to comment on the $1/m^2_Q$ correction effects.
In the above discussion we neglect these effects.
However, if more exact analysis is required,
we should consider these effects.
Because, for a $c$ quark baryon mass,
the magnitude of the effect due to $1/m^2_Q$ corrections is 
about $4\% $ of the baryon mass. 

In this paper, the heavy baryon mass spectra is discussed by taking
into account the ${\sl\Xi}_Q$-${\sl\Xi}'_Q$ mixing from $1/m^{}_Q$ order 
corrections.
By use of the sum rule (\ref{eqn:angle}) we estimate the
${\sl\Xi}_Q$-${\sl\Xi}'_Q$ mixing angle,
$\cos ^2\theta =0.96\pm 0.23$,
in the conventional assignment for charmed baryons.
This is very small if we employ this central value.
In this assignment, the Gell-Mann$-$Okubo mass relation
is effective.
This mass relation gives $\cos ^2\theta =0.88\pm 0.12$.
The origin of the large uncertainty of these two values is almost
entirely the experimental error of the ${\sl\Xi}'_Q$ mass.
On the other hand, Falk's new assignment
leads to the allowed ${\sl\Sigma}_c$ mass
region, ${\sl\Sigma}_c > 2372$MeV.
We are eager for an experimental measurement of the total spin-parity
of ${\sl\Sigma}_c(2453)$ in order to clarify the baryon assignment.

\newpage

{\bf Acknowledgment}

\vspace{1em}

We are grateful to Professor T. Matsuoka for useful advice.


\vspace{2em}

\begin{thebibliography}{99}
\bibitem{rf:1}
N. Isgur and M.B. Wise, Phys. Lett. {\bf B232} (1989), 113;
{\bf B237} (1990), 527.
\bibitem{rf:2}
M. Neubert, Phys. Rep. {\bf 245} (1994), 259.
\bibitem{rf:5}
T. Ito, T. Morii and M. Tanimoto, Z. Phys. {\bf C59} (1993), 57.
\bibitem{rf:6}
N. Isgur and M.B. Wise, Phys. Rev. Lett. {\bf 66} (1991), 1130.
\bibitem{rf:7}
A. F. Falk, JHU-TIPAC-96007, hep-ph/9603389.
\bibitem{rf:8}
M. E. Luke, Phys. Lett. {\bf B252} (1990), 447.
\bibitem{rf:9}
Particle Data Group, Phys. Rev. {\bf D50} (1994),
 `Review of Particle Properties' \\
and 1995 off-year partial update for the 1996 edition.
\bibitem{rf:10}
WA89 Collaboration, `Charmed Baryon Production in the CREN Hyperon Beam', \\
presented by E. Chudakov at Heavy Quark '94, Virginia, October, 1994.
\bibitem{rf:11}
CLEO Collaboration, Phys. Rev. Lett. {\bf 75} (1995), 4364.
\bibitem{rf:12}
DELPHI Collaboration, DELPHI 95-107.
\end{thebibliography}
\end{document}